\documentstyle[12pt,epsfig]{article}
\begin{document}
\vspace{1.0in}
\begin{center}
\vspace{1cm}{\large\bf Dynamics of cascade three level
system interacting \\with the classical and quantized field}\\
\vspace{1cm}{\bf Mihir Ranjan Nath {\footnote[1]
{mrnath\_95@rediffmail.com}} and Surajit Sen {\footnote[2]
{ssen55@yahoo.com}}
\\Department of
Physics\\Guru Charan College\\Silchar 788004\\
Assam, INDIA\\} \vspace {1cm} {\bf Gautam Gangopadhyay
\footnote[3]
{gautam@bose.res.in}\\
 S N Bose National Centre for Basic Sciences\\ JD Block, Sector III \\ Salt Lake City \\
Kolkata-700098, INDIA} \\ \vspace {.5cm}
\end{center}
\begin{center}
\large ({\it Pramana-Journal of Physics $\mathbf{61}$ (2003) 1089})
\end{center}
\begin{abstract}
We study the exact solutions of the cascade three-level atom
interacting with a single mode  classical and quantized field with
different initial conditions of the atom. For the semiclassical
model, it is found that if the atom is initially in the middle
level, the time dependent populations of the upper and lower levels
are always equal. This dynamical symmetry exhibited by the classical
field is spoiled on quantization of the field mode. To reveal  this
nonclassical effect an Euler matrix formalism is developed to solve
the dressed states of the cascade Jaynes-Cummings model (JCM).
Possible modification of such effect on the collapse and revival
phenomenon is also discussed by taking the quantized field in a
coherent state.
\end{abstract}
\begin{flushleft}
\end{flushleft}
\begin{center}
\large I.Introduction
\end{center}
\par
Over the decades, studies of the population inversion of the two,
three and multilevel system have been proved to be an important
tool to understand various fundamental aspects of quantum optics
[1,2]. Many interesting coherent phenomena are observed if the
number of involved levels exceeds two. In particular, the three
level system exhibits a rich class of coherent phenomena such as
two photon coherence [3], double resonance process [4], three
level super-radiance [5], coherent multistep photo-ionization [6],
trilevel echoes [7], STIRAP [8], resonance fluorescence [9],
quantum jump [10], quantum zeno effect [11] etc [12-16]. From
these studies, it is intuitively clear that the atomic initial
conditions of the three level system can generate diverse quantum
optical effects which are not usually displayed by a two level
system [17-20]. The idea of the present investigation is to
enunciate the three level system for various initial conditions
while taking the  field mode to be either classical or
quantized. In this paper the three level system is modelled by the
matrices which are spin-one representation of SU(2) group. A
dressed-atom approach is developed where the Euler matrix is used
to construct the dressed states. We discuss the time development
of the probabilities both for the semiclassical model and the
cascade JCM for various initial conditions and point out the
crucial changes. Finally the collapse and revival phenomenon is
presented taking the quantized field initially in a coherent state.
\par
The subsequent Sections of the paper are organized as follows. To
put our treatment in proper perspective, in Section II we have
derived the probabilities of three levels taking the
field as a classical field. The cascade JCM and its solution in
the rotating wave approximation (RWA) is presented in Section III.
In section IV we have numerically analyzed the time dependent
atomic populations and compared with the semiclassical situation by
taking the quantized field initially in a number state and
in a coherent state. Finally, in conclusion, we
highlight the outcome of our paper and make some pertinent
remarks.
\begin{center}
\large II.The Semi-classical Model
\end{center}

\par
The Hamiltonian to describe the semicalssical problem of a cascade three
level system interacting with a single mode classical field is
\begin{flushleft}
\hspace{1.5in} ${\mathcal{H}}= {\hbar\omega_0}I_{z}
+ \frac{\hbar \omega_1}{\sqrt{2}}(
I_{+} e^{-i\omega t}+I_{-} e^{i\omega t}),$\hfill(1)
\end{flushleft}
where $I$'s represent the spin-one representation of $SU(2)$ matrices
corresponding to the cascade three level system with equal energy gaps($\hbar
\omega_0$) between the states, namely,

\begin{flushleft}
\hspace {1.5in} $I_+=\left[ {\begin{array}{*{10}c}
   0 & 1 & 0  \\
   0 & 0 & 1  \\
   0 & 0 & 0  \\
\end{array}} \right]$,\hfill (2a)
\end{flushleft}
\begin{flushleft}
\hspace{1.5in}
 $I_-=\left[ {\begin{array}{*{10}c}
   0 & 0 & 0  \\
   1 & 0 & 0  \\
   0 & 1 & 0  \\
\end{array}} \right]$,\hfill(2b)
\end{flushleft}
\begin{flushleft}
\hspace{1.5in} $I_z=\left[ {\begin{array}{*{10}c}
   1 & 0 & 0  \\
   0 & 0 & 0  \\
   0 & 0 & -1  \\
\end{array}} \right]$.\hfill(2c)
\end{flushleft}
$\hbar \omega_1$ is the interaction energy between the three level
system with the classical field mode of frequency $\omega$ in RWA.
Let the solution of the Schr\"{o}dinger equation,
\begin{flushleft}
\hspace{1.5in} $i\hbar [ \frac{\partial \psi}{{\partial t}} ] =
\mathcal{H}\psi$,\hfill(3)
\end{flushleft}
 with Hamiltonian
(1) is given by
\begin{flushleft}
\hspace{1.5in} $\psi(t)=C_{+}(t)\mid{+}> +\quad
C_{0}(t)\mid{0}>+\quad C_{-}(t)\mid{-}>$,\hfill(4)
\end{flushleft}
where $C_{+}(t)$, $C_{0}(t)$ and $C_{-}(t)$ are the time dependent
normalized amplitudes with the eigen functions given by
\begin{flushleft}
\hspace{1.5in} $\mid{+}>  = \left[ {\begin{array}{*{10}c}
   1  \\
   0  \\
   0  \\
\end{array}} \right]$,
\hfill(5a)
\end{flushleft}
\begin{flushleft}
\hspace{1.5in} $ \mid{0}>  = \left[ {\begin{array}{*{10}c}
   0  \\
   1  \\
   0  \\
\end{array}} \right]$,
\hfill(5b)
\end{flushleft}
\begin{flushleft}
\hspace{1.5in} $ \mid{-}>   = \left[ {\begin{array}{*{10}c}
   0  \\
   0  \\
   1  \\
\end{array}} \right]$.
\hfill(5c)
\end{flushleft}
We now proceed to calculate the probability amplitudes of the
three states. Substituting Eq.(4) into Eq.(3) and equating the
coefficients of $\mid{+}>$, $\mid{0}>$ and $\mid{-}>$ from both
sides we obtain
\begin{flushleft}
\hspace{.4in} $i \hbar\mathop {C_+}\limits^\cdot(t) = \omega_0
C_+(t)+\frac{\omega_1}{\sqrt 2 }\exp(-i\omega t)C_0 (t)$ \hfill(6a)
\end{flushleft}
\begin{flushleft}
\hspace{.4in} $ i\hbar \mathop {C_0 }\limits^ \cdot  (t) =
\frac{\omega_1}{\sqrt 2 } \exp (i\omega t)C_ + (t) +
\frac{\omega_1}{\sqrt 2 } \exp(-i\omega t)C_-(t)\hfill(6b)$
\end{flushleft}
\begin{flushleft}
\hspace{.4in} $ i\hbar \mathop {C_ -  }\limits^ \cdot  (t) =
\frac{\omega_1}{\sqrt 2 } \exp (i\omega t)C_0 (t) -
\omega_0 C_ -  (t),$\hfill(6c)
\end{flushleft}
where the dot represents the derivative with respect to time.

\par
Let the solutions of Eqs.(6) are of the following form,
\begin{flushleft}
\hspace{1.5in} $C_+(t)=A_+\exp(is_+t)$, \hfill(7a)
\end{flushleft}
\begin{flushleft}
\hspace{1.5in} $C_0(t)=A_0\exp(is_0t)$, \hfill(7b)
\end{flushleft}
\begin{flushleft}
\hspace{1.5in} $C_-(t)=A_-\exp(is_-t)$,\hfill(7c)
\end{flushleft}
where A's are the time-independent constants to be determined.
Plucking back Eqs.(7) in Eqs.(6) we obtain
\begin{flushleft}
\hspace{1.5in} $(s_0-\omega+\omega_0)A_{+}+\frac{1}{{\sqrt 2
}}\omega_1A_0=0$ \hfill(8a)
\end{flushleft}
\begin{flushleft}
\hspace{1.5in} $s_0A_0+\frac{1}{{\sqrt 2 }}\omega(A_++A_-)=0$
\hfill(8b)
\end{flushleft}
\begin{flushleft}
\hspace{1.5in} $(s_0+\omega-\omega_0)A_-+\frac{1}{{\sqrt 2
}}\omega_1A_0=0$. \hfill(8c)
\end{flushleft}

 In deriving Eqs.(8), the time independence of the amplitudes $A_+$,
$A_-$ and $A_0$ is ensured by invoking the conditions
$s_+=s_0-\omega$ and $s_-=s_0+\omega$. The solution of Eqs.(8)
readily yields
\begin{flushleft}
\hspace{1.5in} $s_0=0$\hfill(9a)
\end{flushleft}
\begin{flushleft}
\hspace{1.5in} $s_0=\pm\surd[(\omega-\omega_0)^2+\omega_1^2]
 (\equiv\pm\Omega)$ \hfill(9b)
\end{flushleft}
and we have three values of $s_+$ and $s_-$, namely,
\begin{flushleft}
\hspace{1.5in} $s^1_+=-\omega, \quad s^2_+=\Omega-\omega, \quad
s^3_+=-\Omega-\omega$ \hfill(10a)
\end{flushleft}
\begin{flushleft}
\hspace{1.5in} $s^1_-= \omega,\qquad s^2_-=\Omega+\omega,\quad
s^3_-=-\Omega+\omega$. \hfill(10b)
\end{flushleft}
Using Eqs.(10), Eqs.(7) can be written as
\begin{flushleft}
\hspace{1.2in} $C_+(t)=A_+^1\exp[-i\omega
t]+A_+^2\exp[i(\Omega-\omega)t]+A_+^3\exp[i(-\Omega-\omega)t]\hfill(11a)$
\end{flushleft}
\begin{flushleft}
\hspace{1.2in} $C_0 (t) = A_0^1  + A_0^2 \exp (i\Omega t) + A_0^3
\exp (-i\Omega t)$ \hfill(11b)
\end{flushleft}
\begin{flushleft}
\hspace{1.2in} $C_-(t)=A_-^1\exp(i\omega
t)+A_-^2\exp[i(\Omega+\omega)t]+A_-^3\exp[i(-\Omega+\omega)t].$
\hfill(11c)
\end{flushleft}
where As' be the constants to be calculated from the following
initial conditions:
\par
 {\bf Case-I}\quad :\quad Let us consider at $t=0$, the atom is in the
lower level i.e, $C_+(0)=0$, $C_0(0)=0$, $C_-(0)=1$. Using Eqs.(6)
and (11), the time dependent probabilities of the three levels are
given by
\begin{flushleft}
\hspace{1.5in} $ \left| {C_ +  (t)} \right.|^2  = \frac{{\omega _1
^4 }}{{\Omega ^4 }} \sin^4{\Omega t/2},$\hfill(12a)
\end{flushleft}
\begin{flushleft}
\hspace{1.1in}$ |C_0 (t)|^2  = \frac{{\omega _1 ^2 }}{{2\Omega ^4
}}[4 (\omega  - \omega _0 )^2 \sin^4{\Omega t/2}  + \Omega ^2
\sin^2 \Omega t],$\hfill(12b)
\end{flushleft}
\begin{flushleft}
\hspace{0.7in}
 $\left| {C_ -  (t)} \right.|^2  = \frac{1}{{\Omega
^4 }}[( {\omega _1 ^2 }\sin^2{\Omega t/2} + \Omega ^2 \cos{\Omega
t})^2  + (\omega  - \omega _0 )^2 \Omega ^2 sin^2 {\Omega
t}].\hfill(12c)$
\end{flushleft}
\par
{\bf Case-II}\quad :\quad If we choose the atom is initially in the
middle level, i.e, $C_+(0)=0$, $C_0(0)=1$, $C_-(0)=0$,
the  corresponding probabilities of the levels
are given by
\begin{flushleft}\hspace{.5in}$
|C_ +  (t)|^2= \frac{{\omega _1 ^2 }}{{2\Omega ^4 }}[4(\omega  -
\omega _0 )^2 \sin^4{\Omega t/2}+ \Omega ^2 \sin^2 {\Omega t}]$
\end{flushleft}
\begin{flushleft}
\hspace{1.1in}$=|C_-(t)|^2,$\hfill(13a)
\end{flushleft}
\begin{flushleft}\hspace{.5in}
$\left| {C_0 (t)} \right.|^2  = \frac{{4(\omega  - \omega _0 )^4
}}{{\Omega ^4 }}\sin^4{\Omega t/2}$
\end{flushleft}
\begin{flushleft}\hspace{1.1in}
$+ \frac{{4(\omega  - \omega _0 )}}{{\Omega ^2 }}^2 \sin^{2}{\Omega
t/2}\cos{\Omega t} + \cos^2 {\Omega t}.$\hfill(13b)
\end{flushleft}
Here we note that, unlike the previous case, the probabilities of
the upper and lower levels are equal.

\par
{\bf Case-III}\quad :\quad When the atom is initially in the
upper level i.e, $C_+(0)=1$, $C_0(0)=0$, $C_-(0)=0$,
 we obtain the
following occupation  probabilities in the three levels,
\begin{flushleft}
\hspace{.5in}
 $\left| {C_ +  (t)} \right.|^2  = \frac{1}{{\Omega
^4 }}[( \omega _1^2\sin^2{\Omega t/2} + \Omega ^2 \cos{\Omega t})
^2  + (\omega  - \omega _0 )^2 \Omega ^2 \sin^2 {\Omega
t}],\hfill(14a)$
\end{flushleft}
\begin{flushleft}\hspace{1.1in}
$ \left| {C_0 (t)} \right.|^2  = \frac{{\omega _1 ^2 }}{{2\Omega
^4 }}[4(\omega  - \omega _0 )^2 \sin^4{\Omega t/2}  + \Omega ^2
\sin^2 {\Omega t}],$\hfill(14b)
\end{flushleft}
\begin{flushleft}\hspace{1.7in}
$ \left| {C_ -  (t)} \right.|^2  = \frac{{\omega _1 ^4 }}{{\Omega
^4 }}\sin^4{\Omega t/2}$.\hfill(14c)
\end{flushleft}

We note that the probability of the middle level for case-III is
precisely identical to that of case-I while those of the upper and
lower levels are interchanged.

\begin{center}
 \large III. Cascade Jaynes-Cummings Model
 \end{center}
 \par
Here we consider the cascade three level system interacting with a single mode
quantized field.
 The cascade JCM system in the rotating wave approximation [17,18] is described by the
Hamiltonian
\begin{flushleft}\hspace{1.5in}
$ H = \hbar
\omega(a^{\dag}a+I_z)+({\Delta}I_z+g\hbar(I_+a+I_-a^\dag)),
 $\hfill(15)
\end{flushleft}
where $a^\dag$ and $a$ be the creation and annihilation operators,
$g$ be the coupling constant and ${\Delta}=\hbar(w_{0}-w)$ be the
detuning frequency respectively. It is easy to check that both
diagonal and interaction parts of the Hamiltonian commute with
each other. The eigenfunction of this Hamiltonian is given by
\begin{flushleft}
\hspace{1.5cm}$\mathop {|\psi}\nolimits_{n}(t)  >  =
\sum\limits_{n = 0}^\infty [\mathop C\nolimits_{-}^{n+1}
(t)|n+1,->  + \mathop C\nolimits_{0}^n (t)|n,0>  + \mathop
C\nolimits_{+}^{n-1} (t)|n-1,+> ].$\hfill(16)\\
\end{flushleft}
We note that the Hamiltonian couples the atom-field states
$|n-1,+>$, $|n,0>$ and $|n+1,->$, where n represents
the number of photons of the field.
The interaction part of the
Hamiltonian (15) can also be written in matrix form
\begin{flushleft}
\hspace{1.5in}$H_{int} = \left[ {\begin{array}{*{20}c}
   {-\Delta}  & {g{\hbar}\sqrt {n + 1} } & 0  \\
   {g{\hbar}\sqrt {n + 1} } & 0 & {g{\hbar}\sqrt n }  \\
   0 & {g{\hbar}\sqrt n } &  \Delta   \\
\end{array}} \right].
 $\hfill(17)
 \end{flushleft}
At resonance ($\Delta= 0$), the eigenvalues of the Hamiltonian are
given by $\lambda_+=g{\hbar}\sqrt{2n + 1}$, $\lambda _0 = 0 $ and
$\lambda _-=-g{\hbar}\sqrt{2n + 1}$ with
the corresponding dressed eigenstates
\begin{flushleft}
\hspace{1.5in}$\left[ {\begin{array}{*{20}c}
   {|n,1> }  \\
   {|n,2> }  \\
   {|n,3> }  \\
\end{array}} \right] = T\left[ {\begin{array}{*{20}c}
   {|n+1,-> }  \\
   {|n,0> }  \\
   {|n-1,+> }  \\
\end{array}} \right].$\hfill(18)
\end{flushleft}
In Eq.(18), the dressed states are constructed by rotating the
bare states with the Euler matrix $T$ parameterized as
\begin{flushleft}
\hspace{1.5in} $T = \left[ {\begin{array}{*{20}c}
   {\alpha _{11} } & {\alpha _{12} } & {\alpha _{13} }  \\
   {\alpha _{21} } & {\alpha _{22} } & {\alpha _{23} }  \\
   {\alpha _{31} } & {\alpha _{32} } & {\alpha _{33} }  \\
\end{array}} \right],$\hfill(19)
\end{flushleft}
where
\begin{flushleft}
\hspace{1.5in}
 $\begin{array}{l}
 \alpha _{11}  = cos\psi cos\phi  - cos\theta sin\phi sin\psi  \\
 \alpha _{12}  = cos\psi sin\phi  + cos\theta cos\phi sin\psi  \\
 \alpha _{13}  = sin\psi sin\theta  \\
 \alpha _{21}  =  - sin\psi cos\phi  - cos\theta sin\phi cos\psi  \\
 \alpha _{22}  =  - sin\psi sin\phi  + cos\theta cos\phi cos\psi  \\
 \alpha _{23}  = cos\psi sin\theta  \\
 \alpha _{31}  = sin\theta sin\phi  \\
 \alpha _{32}  =  - sin\theta cos\phi  \\
 \alpha _{33}  = cos\theta. \\
 \end{array}$
\end{flushleft}
The evaluation of its various elements is presented in the
appendix and here we quote the results as follows
\begin{flushleft}
\hspace{1.5in}$\begin{array}{l}
 \alpha _{11}  = \sqrt {\frac{{n + 1}}{{4n + 2}}} , \alpha _{12}  = \frac{1}{{\sqrt 2 }},
 \alpha _{13}  = \sqrt {\frac{n}{{4n + 2}}},\\
 \alpha _{21}  =  - \sqrt {\frac{n}{{2n + 1}}}, \alpha _{22}  = 0,
\alpha _{23}  = \sqrt {\frac{{n + 1}}{{2n + 1}}},\\
 \alpha _{31}  =\sqrt {\frac{{n + 1}}{{4n + 2}}},
\alpha _{32}  =  - \frac{1}{{\sqrt 2 }} ,
\alpha _{33}  = \sqrt {\frac{n}{{4n + 2}}}.
\end{array}$\hfill(20)\\
\end{flushleft}
 The time dependent probability amplitudes of the three
 levels are given by
\begin{flushleft}
\hspace{1.5in} $\left[ {\begin{array}{*{20}c}
   {C_{-}^{n+1} (t)}  \\
   {C_{0}^n (t)}  \\
   {C_{+}^{n-1} (t)}  \\
\end{array}} \right] = T^{ - 1} \left[ {\begin{array}{*{20}c}
   {e^{ - i\Omega_n t} } & 0 & 0  \\
   0 & 1 & 0  \\
   0 & 0 & {e^{i\Omega_n t} }  \\
\end{array}} \right]T\left[ {\begin{array}{*{20}c}
   {C_{-}^{n+1} (0)}  \\
   {C_{0}^n (0)}  \\
   {C_{+}^{n-1} (0)}  \\
\end{array}} \right],$\hfill(21)
\end{flushleft}
where $\Omega_n= g\sqrt {2n + 1}$. In the following we consider different
initial condition of the atom with the quantized field in a number state
$|n>$.

{\bf Case-IV}: Here we consider the atom is initially polarized in the
lower level and the combined atom-field state is
$|n+1,->$, i.e,
$C_{+}^{n-1} (0)=0$, $C_{0}^n(0)=0$, $C_{-}^{n+1}(0) = 1$.
 Using Eqs(20) and (21) the time dependent
atomic population of the three levels are given by
\begin{flushleft}\hspace{1.5in}
$\left| {C_{+}^{n-1} (t)} \right|^2  = \frac{{4n(n + 1)}}{{(2n +
1)^2 }}sin^4 {\Omega_n} t/2,$\hfill(22a)
\end{flushleft}
\begin{flushleft}\hspace{1.5in}
$\left| {C_{0}^n (t)} \right|^2  = \frac{{(n + 1)}}{{(2n + 1)}}
sin^2 {\Omega_n} t,$\hfill(22b)
\end{flushleft}
\begin{flushleft}\hspace{1.5in}
$\left| {C_{-}^{n+1} (t)} \right|^2 = 1 - 4[\frac{{n(n + 1)}}{{(2n
+ 1)^2 }} + \frac{{(n + 1)^2 }}{{(2n + 1)^2 }}cos^2 {\Omega_n}
t/2] sin^2 {\Omega_n} t/2.$\hfill(22c)
\end{flushleft}
{\bf Case-V}: At $ t=0 $ when the atom is in the middle level and the
combined atom-field state
is $|n,0>$ i.e, $C_{+}^{n-1} (0)=0$, $C_{0}^n (0)=1$, $C_{-}^{n+1}(0)=0$, we
find
\begin{flushleft}\hspace{1.5in}
$\left| {C_{+}^{n-1} (t)} \right|^2  = \frac{n}{{(2n + 1)}}sin^2
{\Omega_n} t,$\hfill(23a)
\end{flushleft}
\begin{flushleft}\hspace{1.5in}
$\left| {C_{0}^n (t)} \right|^2  = cos^2 {\Omega_n} t$\hfill(23b)
\end{flushleft}
\begin{flushleft}\hspace{1.5in}
$\left| {C_{-}^{n+1} (t)} \right|^2  = \frac{{(n + 1)}}{{(2n +
1)}}sin^2 {\Omega_n} t.$\hfill(23c)
\end{flushleft}
{\bf Case-VI}: $C_{+}^{n-1} (0) = 1$, $C_{0}^n (0) = 0$,
$C_{-}^{n+1} (0) =0$\\
If the atom is initially in the upper level and the atom-field state is
$|n-1,+>$, i.e, $C_{+}^{n-1} (0) = 1$, $C_{0}^n (0) = 0$,
$C_{-}^{n+1} (0) =0$ we
obtain the following probabilities
\begin{flushleft}\hspace{1.5in}
$\left| {C_{+}^{n-1} (t)} \right|^2  = 1 - 4[\frac{{n(n +
1)}}{{(2n + 1)^2 }} + \frac{{n^2 }}{{(2n + 1)^2 }}cos^2 {\Omega_n}
t/2] sin^2 {\Omega_n}t/2,$\hfill(24a)
\end{flushleft}
\begin{flushleft}\hspace{1.5in}
$\left| {C_{0}^n (t)} \right|^2  = \frac{n}{{(2n + 1)}}sin^2
{\Omega_n} t,$\hfill(24b)
\end{flushleft}
\begin{flushleft}\hspace{1.5in}
$\left| {C_{-}^{n+1} (t)} \right|^2  = \frac{{4n(n + 1)}}{{(2n +
1)^2 }}sin^4 {\Omega_n} t/2$.\hfill(24c)
\end{flushleft}
Finally we note that, at resonance, for large value of $n$ the
probabilities of case-IV, V and VI are identical to case-I, II and
III, respectively indicating the validity of the correspondence
principle.
\begin{center}
\large IV. Numerical results
\end{center}
\par
To explore the physical content, we now proceed to analyze the
probabilities of the semiclassical model and the cascade JCM
numerically.
\par
 For the classical field at resonance, the time evolution of the
probabilities $|C_{+}(t)|^2$ (solid line), $|C_{0}(t)|^2$ (dashed
line) and $|C_{-}(t)|^2$ (dotted line) corresponding to Case-I, II
and III, respectively are shown in Fig 1. We note that for the cases with
atom initially in lower and upper level,
which are displayed in Fig 1(a) and 1(c), respectively
 the probabilities $|C_{+}(t)|^2$ and $|C_{-}(t)|^2$ can attain a
maximum value equal to unity while $|C_{0}(t)|^2$ cannot.
If we compare these two figures the time dependent populations of the lower
and upper levels are different by a phase lag corresponding to the initial
condition of population.
This
clearly shows that the probabilities oscillate between the levels
$\mid+>$ and $\mid->$ alternatively at a Rabi frequency of
$\nu_{I}=\frac{\omega_{1}}{2 \pi}$. On the contrary, the plot of
case-II where the atom is initially in the middle level depicted in Fig 1(b)
 shows that the system oscillates with a
Rabi frequency of $\nu_{II}=\frac{\omega_{1}}{\pi}$ such that the
probabilities of $\mid+>$ and $\mid->$ states are always equal. When
the atom is initially in the middle level, the exactly sinusoidal
resonant field interacts with the atom in such a way that the upper
and lower levels are dynamically treated on an equal footing. This
dynamically symmetrical distribution of population between the upper
and lower levels is possible because of the classical field.
\par
 For quantized field  we consider the time evolution of
the probabilities in two different situations of initial condition
of the field: a) when the field is in a number state and b) the
field is in a coherent state.
\par
a) For the cascade JCM, the probabilities of case-IV, V and VI are
plotted in Fig 2 when the field is in a number state with $n=1$ and g=0.1.
In Fig 2(a)
we note that for case-IV i.e, when the atom is initially in the lower level,
 the
Rabi-frequency of oscillation is $\nu_I^n =
\frac{\Omega_n}{2\pi}$. However, unlike case-I of the
semiclassical model, the probabilities $|C_{+}^{n-1}(t)|^2$ never
becomes unity.
On the other hand, Fig 2(b) illustrates the probabilities of case-V i.e, when
the atom is initially in the middle level,
where the system oscillates with a Rabi frequency
$\nu_{II}^n=\frac{\Omega_n}{\pi}$ and once again, in contrast to the
corresponding semiclassical situation in
case-II, the probabilities of the upper and lower level are not
equal. The probabilities of case-VI i.e, when the atom is initially in the
upper level, depicted in Fig 2(c) shows that
although it possesses the same Rabi frequency $\nu_I^n$, but the
pattern of oscillation is not out of phase of case-IV.
To compare with one can look back the semiclassical interaction where
we have shown that in case-I the
pattern of oscillation of upper(lower) level population is precisely
identical to the lower(upper) level population of case-III.
\par
To understand the implications of such dynamical symmetry
breaking qualitatively, various bounds on the probabilities are
given below :
\begin{center}
Table-I
\end{center}
\begin{center}
\begin{tabular}{|c|c|c|c|}
  \hline
  Case & Semiclassical model & Case & Cascade JCM \\
  \hline
    & $0\leq|C_{-}(t)|^2\leq 1$, &  & $0\leq|C_{-}^n(t)|^2\leq 1$, \\
  I & $0\leq|C_{0}(t)|^2< 1$, & IV & $0\leq|C_{0}^n(t)|^2<1$,  \\
    & $0\leq|C_{+}(t)|^2\leq 1$ &   & $0\leq |C_{+}^n(t)|^2<1$ \\
  \hline
    & $0\leq|C_{-}(t)|^2< 1$, &   & $0\leq|C_{-}^n(t)|^2<1$, \\
  II & $0\leq|C_{0}(t)|^2\leq 1$, & V & $0\leq|C_{0}^n(t)|^2\leq 1$, \\
    & $0\leq|C_{+}(t)|^2< 1$ &  & $0\leq |C_{+}^n(t)|^2<1$ \\
  \hline
    & $0\leq|C_{-}(t)|^2\leq 1$, &  & $0\leq |C_{-}^n(t)|^2<1$,  \\
   III & $0\leq|C_{0}(t)|^2< 1$, & VI & $0\leq |C_{0}(t)^n|^2<1$, \\
   &  $0\leq|C_{+}(t)|^2\leq 1$  &  & $0<|C_{-}^n(t)|^2\leq 1$ \\ \hline
\end{tabular}
\end{center}
We note that for the semiclassical model, the symmetric evolution
of the probabilities results an identical bounds for case-I and
III as shown in the Table-I. On quantization of the field mode,
the bounds corresponding to case-IV and VI are no longer similar
although those for case-II remains same as case-V. At resonance,
for large value of $n$, Eqs.(22), (23) and (24) of case-IV, V and
VI are precisely identical to Eqs.(12), (13) and (14) of case-I,
II and III, respectively and we recover the same bounds of the
semiclassical model.
\par
b) Finally, we consider the atom is interacting with the quantized field mode
 in a coherent state. The coherently averaged probabilities of case-IV, V
and VI are given by
\begin{flushleft}\hspace{1.2in}
$<P_{ +  }(t)> = \sum\limits_n {P_{n} \left| {C_{+}^{n-1} }(t)
\right|^2 } $\hfill(25a)\end{flushleft}
\begin{flushleft}\hspace{1.2in}
$<P_{  0}(t)> =\sum\limits_n {P_{n}\left| {C_{0}^{n} }(t)
\right|^2  }$\hfill(25b)
\end{flushleft}
\begin{flushleft}\hspace{1.2in}
$<P_{ -}(t)>=\sum\limits_n {P_{n}\left| {C_{-}^{n+1} }(t)
\right|^2 }$,\hfill(25c)
\end{flushleft}
where $P_{n} = \frac{exp[-\bar{n}]\bar{n}^n}{n!}$ be the Poisson
distribution function and $\bar{n}$ be the mean photon number. For
all numerical purpose we choose $g=0.1$. We have studied
extensively for various values of $\bar{n}$. The figures are given
only for $\bar{n}=50$. The Figs 3-5 display the numerical plots of
Eqs.(25) where the collapse and revival of the Rabi oscillation is
clearly evident. For low $\bar{n}$ and when the atom is in the
middle level the symmetrical values of population of upper and
lower levels are not observed until $\bar{n}$ is very high as
given in the figures. However, even if $\bar{n}=50$, the numerical
values of the time dependent populations of the upper and lower
levels are not exactly equal although very close and becomes
exactly equal in the limit $\bar{n}\rightarrow \infty$. We further
note that, if the atom is initially polarized either in the upper
or in the lower level, it exhibits similar population oscillation,
which is different from the case if it is initially polarized in
the middle level. The reproduction of this result analogous to the
semiclassical model shows the proximity of the coherent state with
large $\bar{n}$ to the classical field.
\par
When the field is quantized, the population oscillation depends on
the occupation number, $n$ of the field state, for example,
$\cos{(g\sqrt{2n+1}t)}$. For a statistical distribution of field
state, the spontaneous factor $1$ plays a dominant role when $n$
is low. For an initial number state of the field when $n$ is
slightly higher than $1$, the upper and lower levels of the atom
are not treated dynamically on an equal footing even when the atom
is initially in the middle level. This fine graining of the
quantized distribution of photons over the number states $\{|n>\}$
generates a complex interference between individual Rabi
oscillations corresponding to each $n$ and plays a role until when
$n$ is very large compared to $1$ and effectively acts as a
classical field and thereby the semiclassical situation is
satisfied. Note that for an initial vacuum field, i.e, $n=0$ for
the number state and $\bar{n} =0$ for the coherent state, with the
atom initially in the middle level, it can not go to the upper
level at all and the population will oscillates between the lower
and middle levels with Rabi frequency $\Omega_0$. This asymmetry
is still present when the field is in a coherent state with a
Poissonian  photon distribution with low average photon number,
$\bar{n}$, which is generally not symmetric around an $\bar{n}$. A
Poisson distribution is almost symmetric, a Gaussian, around a
$\bar{n}$ if $\bar{n}$ is very large which is the case of a
classical field. In that situation the upper and lower levels of
the atom are treated dynamically on an equal footing and maintains
the symmetrical distribution of population in upper and lower
levels.
\begin{center}
\large V. Conclusion
\end{center}
\par
We conclude by recapitulating the essential content of our
investigation. At the outset we have sculpted the semiclassical
model by choosing the spin-one representation of $SU(2)$ group and
have calculated the transition probabilities of the three levels.
It is shown that at resonance, if the atom is initially polarized
in the lower or in the upper level, the various atomic populations
 oscillate quite differently
when it is initially populated in the middle level. When the atom is
initially populated in the middle level, the classical field
interacts in such a way that the populations of the upper and lower
levels are always equal. This dynamically symmetrical populations of
the upper and lower levels are destroyed due to the quantization of
the field. To show this quantum behaviour a dressed-atom approach is
presented to solve the cascade JCM. Finally we discuss the
restoration of the symmetry taking the quantized field in a coherent
state with large average photon number, a closest state to the
classical state. Although the collapse and revival and some other
nonclassical features are well studied in the context of two-level
systems, the above dynamical breaking of symmetry due to the
quantization of the field has no two-level analog. We hope that this
dynamical behaviour in the cascade three level system should show
its signature on the time dependent profile of the second order
coherence of the quantized field which will be discussed elsewhere.
The dressed-atom approach developed here may also find its
application in the $V$ and $\Lambda$ type three level systems where
the nature of the symmetry should be different from the cascade
system. \vfill
\pagebreak
\par
\begin{center}\large APPENDIX \end{center}
At resonance, the interaction part of the Hamiltonian of the three
level system is given by
\begin{flushleft}
\hspace{1.5in} $H_{int} = \left[{\begin{array}{*{20}c}
   0 & {g\sqrt {n + 1} } & 0  \\
   {g\sqrt {n + 1} } & 0 & {g\sqrt n }  \\
   0 & {g\sqrt n } & 0  \\
\end{array}} \right],$\hfill(A.1)
\end{flushleft}
where the eigenvalues are $ \lambda _ +   = g\sqrt {2n + 1}$, $
\lambda _0  = 0$ and $\lambda _ -   =  - g\sqrt {2n + 1}$.
The Euler matrix T,  diagonalizes the
Hamiltonian as $H_D=TH_{int}T^{-1}$, is given by Eq.(19). Using
the trick $(H_{int}-\lambda_{j}I)\{X_{j}\}=0$, where $\{X_{j}\}$
be the column matrix of $T^{-1}$, corresponding to the eigenvalue
$\lambda_+$ we have
\begin{flushleft}
 \hspace{1.1in}$\left[ {\begin{array}{*{20}c}
   { - g\sqrt {2n + 1} } & {g\sqrt {n + 1} } & 0  \\
   {g\sqrt {n + 1} } & { - g\sqrt {2n + 1} } & {g\sqrt n }  \\
   0 & {g\sqrt n } & { - g\sqrt {2n + 1} }  \\
\end{array}} \right]
\left[ {\begin{array}{*{20}c}
   {\alpha_{11}}  \\
   {\alpha_{12}}  \\
   {\alpha_{13}}  \\
\end{array}} \right] = 0.$\hfill(A.2)
\end{flushleft}
These linear equations readily yield
\begin{flushleft}
\hspace{1.5in} $\alpha_{12} = \frac{{\sqrt {2n + 1} }}{{\sqrt n
}}\alpha_{13}$, $\alpha_{12}  = \frac{{\sqrt {2n + 1} }}{{\sqrt {n
+ 1} }}\alpha_{11}$ and $\alpha_{11}  = \frac{{\sqrt {n + 1}
}}{{\sqrt n }}\alpha_{13}$.\hfill(A.3)
\end{flushleft}
Using the normalization condition
\begin{flushleft}
 \hspace{1.5in}$\alpha_{11}^2+\alpha_{12}^2+\alpha_{13}^2 =1$,\hfill(A.4)
\end{flushleft}
we get $\alpha_{11}  = \sqrt {\frac{{n + 1}}{{4n + 2}}}$ ,
$\alpha_{12} = \sqrt {\frac{{2n + 1}}{{4n +
2}}}=\frac{1}{\sqrt{2}}$ and
$\alpha_{13}  = \sqrt {\frac{n}{{4n + 2}}}$.
Similarly, corresponding to the eigenvalues $\lambda_0$ and
$\lambda_-$ we can obtain other elements of $T$, namely,
\begin{flushleft}
\hspace{1.5in}
 $\alpha _{21}  =  - \sqrt {\frac{n}{{2n + 1}}}$ , $\alpha _{22}  = 0$ and $\alpha _{23}  = \sqrt {\frac{{n + 1}}{{2n + 1}}}$, \hfill(A.5)\\
\end{flushleft}
\begin{flushleft}
\hspace{1.5in}
 $\alpha _{31}  =\sqrt {\frac{{n + 1}}{{4n + 2}}}$, $\alpha _{32}  =  - \frac{1}{{\sqrt 2 }}$ and $\alpha _{33}  = \sqrt {\frac{n}{{4n +
 2}}}$,
 \hfill(A.6)
\end{flushleft}
One can now easily read off the Euler's angles
\begin{flushleft}
\hspace{1.5in}
 $sin\theta  = \sqrt {\frac{{3n + 2}}{{4n + 2}}}$,
 $sin\phi  = \sqrt {\frac{n+1}{{3n + 2}}}$ and
 $sin\psi  = \sqrt {\frac{{n }}{{3n + 2}}}$.
\hfill(A.7)\\
\end{flushleft}
\vfill
\begin{center}
\large Acknowledgement
\end{center}
\par
The authors are thankful to University Grants Commission, New Delhi
for extending partial support. SS thanks S N Bose National Centre
for Basic Sciences, Kolkata for hospitality. SS is grateful to
Professor W S Hou for inviting him in National Taiwan University,
Taiwan, where part of the work was carried out.

\pagebreak

\bibliographystyle{plain}

\pagebreak
\begin{center}
\large
\end{center}
\begin{figure}
\centerline{ \epsffile{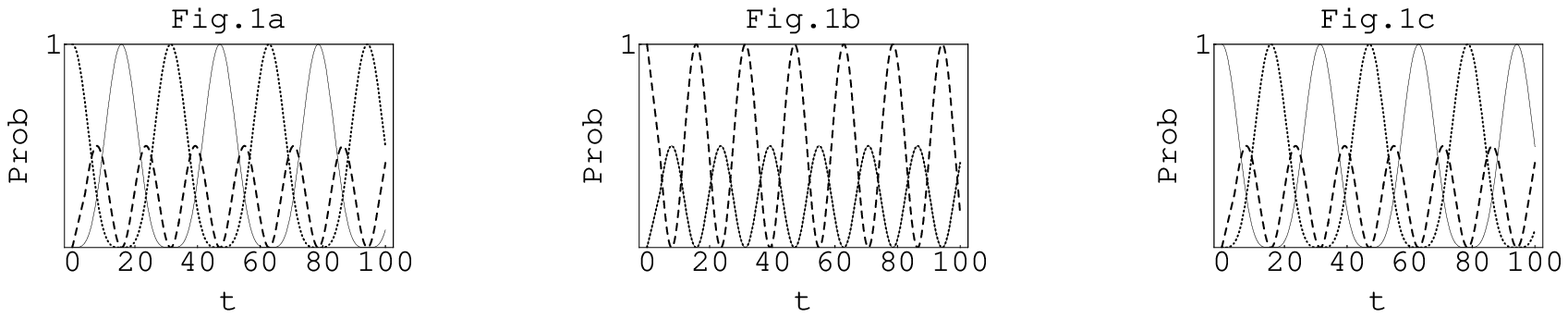}}
\noindent \small \bf {[Fig.1] : The time evolution of the
probabilities of the semiclassical model corresponding to case-I, II
and III. The symmetric pattern of evolution is evident from Fig.1a
and Fig.1c which are in opposite phase.}
\end{figure}
\begin{figure}
\centerline{ \epsffile{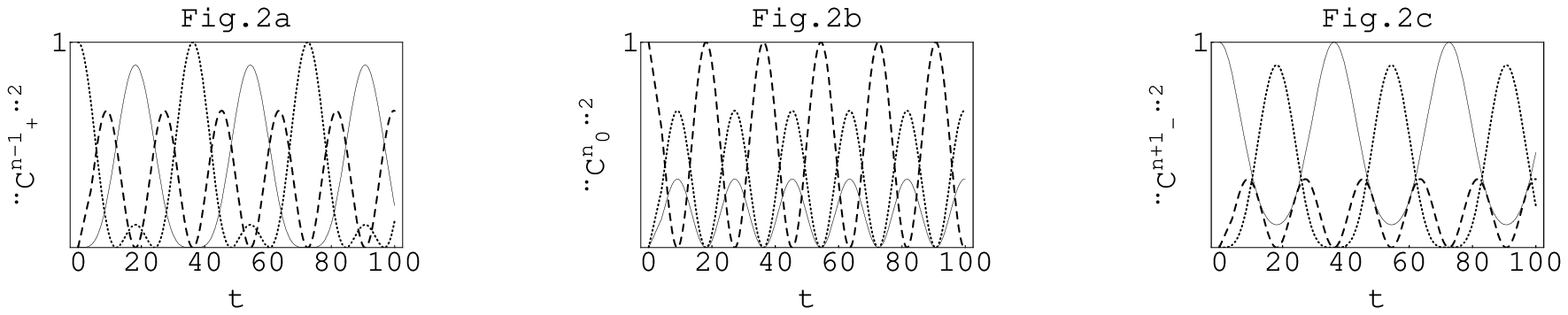}}
\noindent \small \bf {[Fig.2] : The time evolution of the
probabilities of the cascade JCM corresponding to case-IV, V and VI.
In Fig.2a and Fig.2c depict that the symmetry exhibited by the
semiclassical model is spoiled on quantization of the field mode.}
\end{figure}
\pagebreak

\begin{figure}
\centerline{ \epsffile{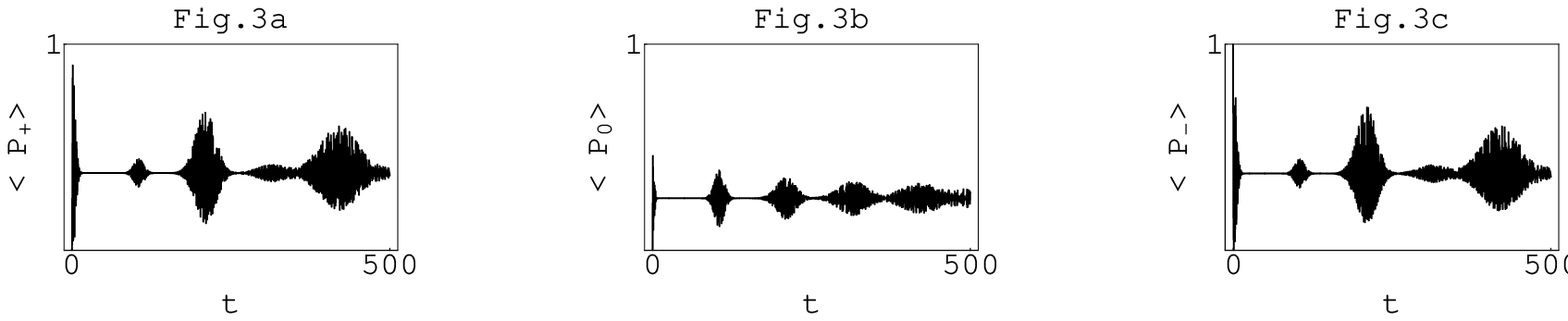}}
\noindent \small \bf {[Fig.3] : The collapse and revival are shown
for case-IV when the atom is initially in the lower level. The time
dependent profiles of the upper and lower level populations are
similar in Fig 3(a) and 3(c).}
\end{figure}
\begin{figure}
\centerline{ \epsffile{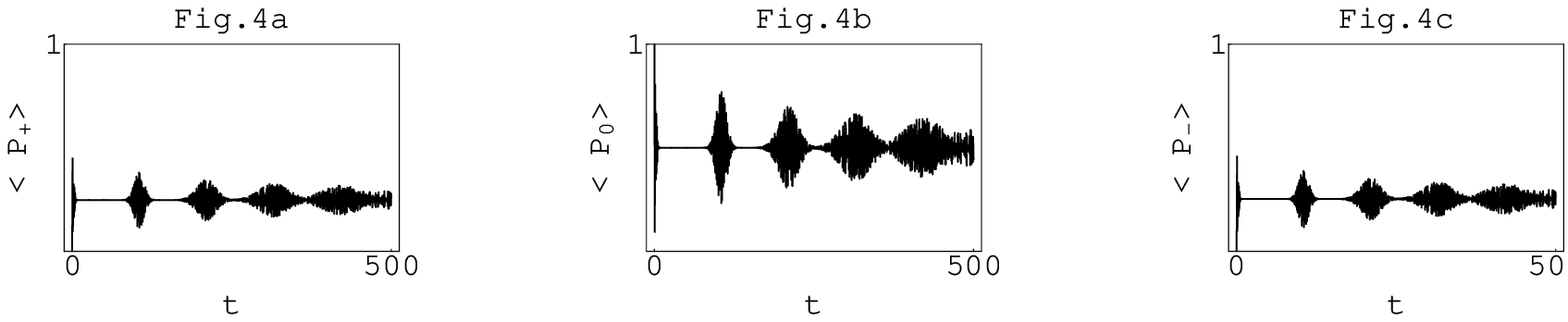}}
\noindent \small \bf {[Fig 4] : The collapse and revival for case-V
are displayed when the atom is initially in the middle level. The
time dependent patterns of the upper and lower level populations are
similar in Fig 4(a) and 4(c) as in the semiclassical cases of Fig
1(b).}
\end{figure}
\begin{figure}
\centerline{ \epsffile{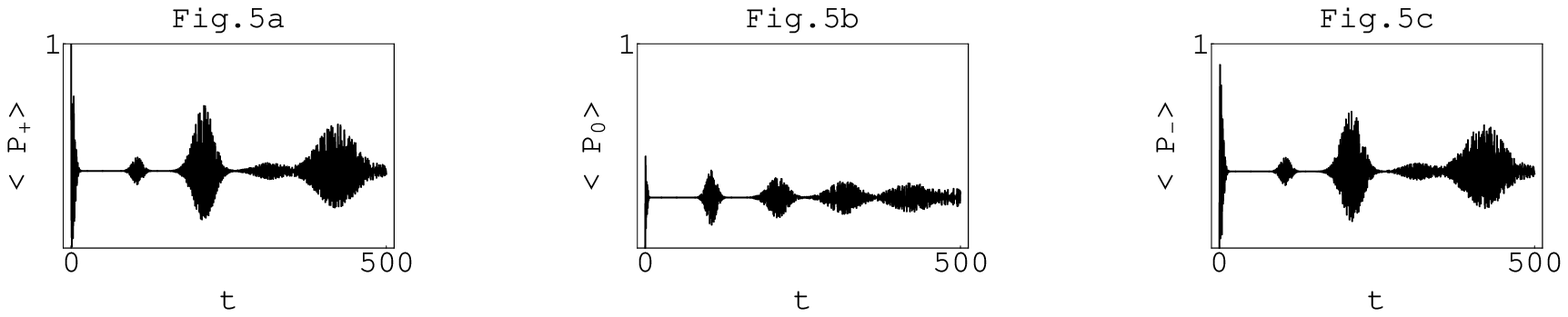}}
\noindent \small \bf {[Fig5] :  The collapse and revival for case-VI
are shown when the atom is initially polarized in the upper level.
Similar to Fig 1(a) and Fig 1(c) of the semiclassical model, Fig 3
and Fig 5 are closely alike.}
\end{figure}
\end{document}